\documentclass[preprint2]{aastex}
\begin{document}

\title{\large{\rm{ANCHORS FOR THE COSMIC DISTANCE SCALE: \\ THE CEPHEID QZ NORMAE IN THE OPEN CLUSTER NGC6067}}}

\author{\small D. Majaess$^1$, L. Sturch$^2$, C. Moni Bidin$^3$, M. Soto$^4$, W. Gieren$^5$, R. Cohen$^5$, F. Mauro$^5$, D. Geisler$^5$, C. Bonatto$^6$, J. Borissova$^7$, D. Minniti$^{8,9,10}$, D. Turner$^{11}$, D. Lane$^{11}$, B. Madore$^{12}$, G. Carraro$^{13}$, L. Berdnikov$^{14}$}

\affil{$^1${\footnotesize Halifax, Nova Scotia, Canada.}}
\affil{$^2${\footnotesize Institute for Astronomical Research, Dept. of Astronomy, Boston University 725 Commonwealth Ave, Boston, MA 02215, USA.}}
\affil{$^3${\footnotesize Instituto de Astronom\'ia, Universidad Cat\'olica del Norte, Av.~Angamos 0610, Antofagasta, Chile.}}
\affil{$^4${\footnotesize Departamento de Astronom\'ia, Universidad de Chile, Casilla 36-D, Santiago, Chile.}}
\affil{$^5${\footnotesize Departamento de Astronom\'ia, Universidad de Concepci\'on, Casilla 160-C, Concepci\'on, Chile.}}
\affil{$^6${\footnotesize Departamento de Astronomia, Universidade Federal do Rio Grande do Sul, Av.~Bento Gonalves 9500 Porto Alegre
91501-970, RS, Brazil.}}
\affil{$^7${\footnotesize Departamento de F\'isica y Astronom\'ia, Facultad de Ciencias, Universidad de Valpara\'iso, Av.~Gran Breta\~na 1111, Valpara\'iso, Chile.}}
\affil{$^8${\footnotesize Departamento de Astronom\'ia y Astrof\'isica, Pontificia Universidad Cat\'olica de Chile, Casilla 306,
Santiago, Chile.}}
\affil{$^9${\footnotesize Departamento de Ciencia Fisicas, Universidad Andres Bello, Santiago, Chile.}}
\affil{$^{10}${\footnotesize Vatican Observatory, V00120, Vatican City State, Italy.}}
\affil{$^{11}${\footnotesize Department of Astronomy and Physics, Saint Mary's University, Halifax, NS B3H 3C3, Canada.}}
\affil{$^{12}${\footnotesize Carnegie Observatories, 813 Santa Barbara Street, Pasadena, CA 91101, USA.}}
\affil{$^{13}${\footnotesize European Southern Observatory, Avda Alonso de Cordova, 3107, Casilla 19001, Santiago, Chile.}}
\affil{$^{14}${\footnotesize Moscow M V Lomonosov State University, Sternberg Astronomical Institute, Moscow 119992, Russia.}}
\email{dmajaess@cygnus.smu.ca}

\begin{abstract}
Cepheids are key to establishing the cosmic distance scale.  Therefore it's important to assess the viability of QZ Nor, V340 Nor, and GU Nor as calibrators for Leavitt's law via their purported membership in the open cluster NGC 6067.  The following suite of evidence confirms that QZ Nor and V340 Nor are members of NGC 6067, whereas GU Nor likely lies in the foreground: (i) existing radial velocities for QZ Nor and V340 Nor agree with that established for the cluster ($-39.4\pm 0.2(\sigma_{\bar{x}}) \pm 1.2 (\sigma )$ km/s) to within 1 km/s, whereas GU Nor exhibits a markedly smaller value; (ii) a steep velocity-distance gradient characterizes the sight-line toward NGC 6067, thus implying that objects sharing common velocities are nearly equidistant; (iii) a radial profile constructed for NGC 6067 indicates that QZ Nor is within the cluster bounds, despite being $20\arcmin$ from the cluster center; (iv) new $BVJH$ photometry for NGC 6067 confirms the cluster lies $d=1.75\pm0.10$ kpc distant, a result that matches Wesenheit distances computed for QZ Nor/V340 Nor using the \citet[][HST parallaxes]{be07} calibration. QZ Nor is a cluster Cepheid that should be employed as a calibrator for the cosmic distance scale.
\end{abstract}
\keywords{open clusters and associations: general, stars: distances, stars: variables: Cepheids}

\section{{\rm \footnotesize INTRODUCTION}}
\begin{figure*}[!t]
\begin{center}
\includegraphics[width=17.2cm]{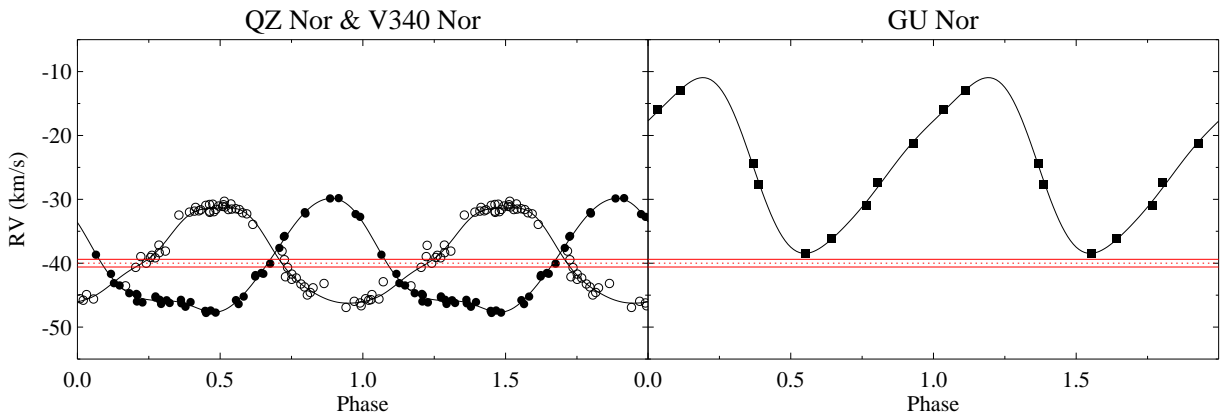} 
\caption{\small{Radial velocity curves for QZ Nor (open circles), V340 Nor (filled circles), and GU Nor (filled squares).  Left panel, the former two Cepheids share common mean velocities that match the cluster velocity ($-39.4\pm 0.2(\sigma_{\bar{x}}) \pm 1.2 (\sigma )$ km/s).  The velocity determined for NGC 6067 is highlighted by the red band, and was determined from 10 cluster members observed by \citet{me08}.  Conversely, GU Nor displays a discrepant velocity (right panel).  Uncertainties associated with the velocity measurements are typically on the order of the symbol size.}}
\label{fig-rvc}
\end{center}
\end{figure*}

Cepheid variables are crucial for defining the cosmic distance scale, determining the Hubble constant ($H_0$), and mitigating degeneracies plaguing the selection of a cosmological model \citep[][for a historical review]{tr12}.  Research on those topics are being led by collaborations such as the Araucaria \citep{gi05,gi06}, S$H_0$ES \citep{mr09}, and Carnegie Hubble projects \citep{fm10}. The latter is the next generation follow-up to the HST key project to measure $H_0$, and aims to reduce existing uncertainties ($10$\%) to less than $3$\%.  The Carnegie Hubble project is relying partly on Cepheids belonging to open clusters to achieve that objective \citep{tu10,mo12}, since the addition of such calibrators reduces uncertainties tied to extragalactic distances computed via Leavitt's law.\footnote{The Cepheid period-luminosity (magnitude) relation.}  Concerns regarding spurious cluster Cepheids are warranted, and the calibration is under constant revision.  However, those concerns can be mitigated by favoring calibrators exhibiting matching cluster and IRSB\footnote{Infrared surface brightness technique \citep{fg97}.} distances \citep{gi13}, unless overwhelming evidence exists supporting the cluster Cepheids in question (e.g., radial velocities and multiband photometry).  

Considerable work remains to bolster the cluster Cepheid calibration, and that includes improving parameters for existing cluster Cepheids and searching for new calibrators \citep{ma12b,an12}. Those objectives motivate the present study, as questions linger concerning QZ Nor, V340 Nor, GU Nor, and their link to NGC 6067.  \citet{wa77} hinted that GU Nor may be a cluster member, but advocated that a reliable cluster distance and set of velocities were ultimately needed to reach a conclusion.  \citet{eg83} argued on the basis of $uvby \beta$ photometry that GU Nor  and V340 Nor feature distances that match stars in Nor OB1, whereas QZ Nor is a member of the foreground NGC 6067.  Conversely, \citet{wa85} and \citet{cc85} asserted that QZ Nor and V340 Nor are members of NGC 6067, and cast doubt on the reported association with GU Nor given the star lies $60\arcmin$ (projected separation) from the cluster center \citep[see also][]{mb86}.    However, new IRSB distances for QZ Nor \citep{st11,gr13} are inconsistent with the distance to NGC 6067, and four recent studies on cluster Cepheids were split on overlooking \citep{an07,tu10} or including QZ Nor as a calibrator \citep{ma08,an12}.  The first iteration of the Carnegie Hubble project likewise bypassed QZ Nor, but V340 Nor was adopted as a calibrator.  V340 Nor lies near the center of NGC 6067, whereas QZ Nor is offset by $20\arcmin$.  That latter separation helped foster concerns regarding the Cepheid's status as a cluster member.

In this study, a multifaceted approach is undertaken to assess the purported link between QZ Nor, V340 Nor, GU Nor, and NGC 6067. First, existing radial velocities for the Cepheids and cluster are examined.  Second, the magnitude of the predicted radial velocity-distance gradient, as inferred from Galactic rotation, is evaluated for the sight-line toward NGC 6067 ($\ell\sim330\degr$).  The objective is to examine whether radial velocities are a reliable indicator of (relative) distance.  In other words, are objects that share common velocities along that sight-line nearly equidistant?  Third, the radial profile for NGC 6067 is mapped using 2MASS\footnote{Two Micron All Sky Survey.} data \citep{cu03} to determine whether QZ Nor lies within the cluster's extent.  Fourth, new $BV$ photometry from the du Pont telescope at Las Campanas Observatory, in concert with new $VVV$\footnote{VISTA Variables in the Via Lactea.} near-infrared $JH$ photometry, are employed to establish a precise independent cluster distance, as sizable offsets exist between the \citet{th62}, \citet{eg83}, and \citet{wa85} distances for NGC 6067.  The $VVV$ survey is a near-infrared campaign sampling part of the 4$^{\rm th}$ Galactic quadrant where NGC 6067 is located \citep{mi10,sa12}.   Lastly, the results are summarized in \S \ref{s-conclusion}.  The analysis will dictate whether QZ Nor, V340 Nor, or GU Nor should be adopted as calibrators to anchor the cosmic distance scale (e.g., for the Araucaria and S$H_0$ES projects).   Such stars are an important means for establishing distances to Local Group galaxies and beyond \citep{in13}, benchmarking standard candles and assessing the impact of compositional differences between target and calibrating stars \citep{mat12,ku12}, precisely defining the galactocentric metallicity and age gradients of the Milky Way and M31 \citep{lu11,ko13}, and constraining the behaviour of intermediate mass stars \citep[][and references therein]{ne12,bo13}.

\section{{\rm \footnotesize ANALYSIS}}
\subsection{{\rm \footnotesize RADIAL VELOCITIES}}
\label{s-rvs}
Radial velocity curves for QZ Nor, V340 Nor, and GU Nor are shown in Fig.~\ref{fig-rvc}.   Radial velocity data for QZ Nor were taken from \citet{cc85}, \citet{me92}, \citet{ki99}, and \citet{gr13}, while for V340 Nor the measurements stem from \citet{me92}, \citet{be94}, and \citet{me08}.  Data for GU Nor were obtained by \citet{me92} and \citet{po94}.  Radial velocity measurements for the Cepheids were phased with an arbitrary ephemeris, and corrections were made to adjust for period changes owing to stellar evolution \citep{tu06,ne12}.  The pulsation periods adopted are from \citet{be00} and \citet{be08}.  Fifth-order Fourier fits were determined for QZ Nor and V340 Nor, whereas a third-order fit was inferred from the sparser dataset characterizing GU Nor.  The Fourier function applied is described by:
$RV={RV}_0+\sum (a_i \cos{2\pi \phi}+b_i \cos{2\pi \phi})$.

Mean radial velocities determined for QZ Nor, V340 Nor, and GU Nor are: ${RV}_0=-40.3\pm0.2,-39.3\pm0.1,-24.72\pm0.01$ km/s.  Formal uncertainties deduced from the Fourier fits for those velocities are misleading owing to potential offsets stemming from inhomogeneous  standardization (different instrumentation and reductions, e.g., CORAVEL cross-correlation), and binarity.  QZ Nor and V340 Nor exhibit $<1$ km/s scatter in the denser sampled regions of the velocity curves, and that value is a more apt indication of the uncertainty. Concerning GU Nor, the phase coverage for the velocity data isn't satisfactory, and the Fourier fit may misrepresent the extrema.  R. Anderson (private communication) obtained $-25.2$ km/s for GU Nor, which is similar to the determination established here. However, additional observations are needed to detect any putative binary companion \citep{sz03} that may affect the deduced velocity for GU Nor \citep[][see the velocity curve for the Cepheid DL Cas]{gi94}.  Alternatively, it has been suggested that the radial velocity offset for GU Nor could indicate that the star was ejected from NGC 6067.

A mean velocity of $-39.4\pm 0.2(\sigma_{\bar{x}}) \pm 1.2 (\sigma )$\footnote{$\sigma_{\bar{x}}$ and $\sigma$ are the standard error and standard deviation, respectively.} km/s was derived for NGC 6067 using 10 members identified by \citet{me08}, and that result is highlighted by the red band in Fig.~\ref{fig-rvc}.  \citet{fm08} obtained an analogous mean velocity, to within the uncertainties.  Fig.~\ref{fig-rvc}, and the aforementioned discussion, imply that QZ Nor and V340 Nor feature velocities that match the cluster, whereas GU Nor is offset.  Lastly, the Cepheid velocities are inconsistent with the velocity established for the planetary nebula (PN, G329.5-02.2) located in the field of NGC 6067, thereby confirming the \citet{mb13} conclusion that the PN is not a cluster member.  

\begin{figure}[!t]
\begin{center}
\includegraphics[width=8.8cm]{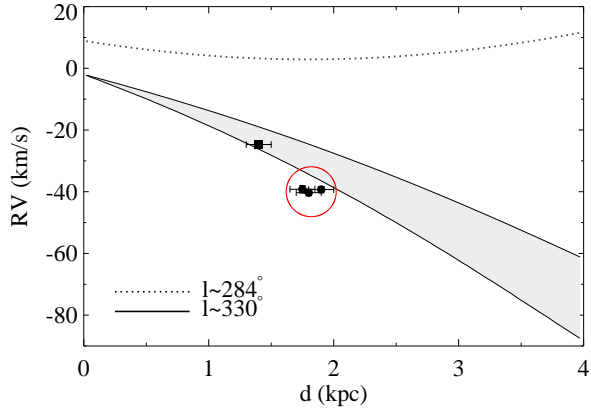} 
\caption{\small{\textit{Approximate} velocity-distance profiles for sight-lines toward NGC 6067 ($\ell\sim330\degr$) and Westerlund 2  ($\ell\sim284\degr$). The sight-line toward NGC 6067 exhibits a steep velocity-distance gradient ($\sim-16\pm3$ km/s/kpc), whereas the opposite is true for Westerlund 2.  Radial velocities for stars in the field of NGC 6067 can be used to estimate \textit{relative} distances.  A spread in potential solutions for the NGC 6067 sight-line is presented, however, shortcomings endemic to the (crude) Galactic rotation model adopted were ignored.  GU Nor (square) is offset from the group encompassed by the red circle, which includes QZ Nor, NGC 6067, and V340 Nor.  The Cepheid distances plotted stem from the Wesenheit function, while the cluster distance is tied to the isochrone fit in Fig.~\ref{fig-cmd}.}}
\label{fig-rvp}
\end{center}
\end{figure}

Radial velocities for younger stellar targets can be a reliable distance indicator along certain sight-lines, provided the objects partake in Galactic rotation: $RV \sim R_0 \sin{\ell}\cos{b}\left( -V_{\sun}/R_0+V_{\sun}/R \right)-9\cos{b}\cos{\ell}-11\cos{b}\sin{\ell}-6\sin{b}$, where $R=\sqrt{{R_0}^2+d^2-2R_0d\cos{\ell}}$. The velocity-distance gradient is rather steep ($\sim-16\pm3$ km/s/kpc) for the line of sight toward NGC 6067 ($\ell\sim330\degr$, Fig.~\ref{fig-rvp}).  Conversely, certain sight-lines, such as toward the cluster Westerlund 2 ($\ell\sim284\degr$), display rather shallow gradients and velocity-distance degeneracies (Fig.~\ref{fig-rvp}).  Hence determining the distance to Westerlund 2 based on kinematic evidence is somewhat more complicated, a fact which is partly responsible for the spread in cited cluster distances \citep[2-8 kpc,][and discussion therein]{ca12}.  For the present analysis the absolute value from the velocity-distance correlation is unimportant, as various uncertainties promulgate into the determination \citep[e.g., distance to the Galactic center and the rotation model adopted,][]{ma10,mal13}.  Rather it is the magnitude of the velocity-distance gradient that is pertinent.  A steep gradient implies that two objects along the sight-line that share common velocities are nearly equidistant.  A pronounced gradient likewise mitigates uncertainties arising from typically imprecise velocity determinations for earlier-type cluster stars associated with Cepheids, since those stars exhibit broad spectral lines (rotation and V-IV luminosity class). The latter isn't relevant here given the cluster (conveniently) hosts numerous red giants \citep{me08}, which were used to compute a precise mean velocity. 

Figs.~\ref{fig-rvc} and \ref{fig-rvp} imply that QZ Nor, V340 Nor, and NGC 6067 are nearly equidistant, whereas GU Nor lies in the foreground.\footnote{The three Cepheids are assumed to follow the velocity-distance correlation, yet there exist stars that do not (e.g., binary interactions, peculiar motions).}  That conclusion is supported by evaluating first-order distances to the Cepheids using existing period-Wesenheit ($VI_c$) relations, which provide a consistency check.  A recent iteration of the Galactic $VI_c$ Wesenheit function is that cited by \citet{ma11b} \citep[see also][]{ng12b}. That iteration is tied principally to the efforts of fellow researchers: the cluster Cepheid calibrators of \citet{tu10}, and the HST parallax Cepheid calibrators of \citet{be07}.  Parameters for the cluster Cepheids TW Nor, SU Cas, $\delta$ Cep, and $\zeta$ Gem have since been revised \citep[e.g.,][]{ma12}, and continued revisions are expected.  Nevertheless, to avoid biasing the analysis, distances are computed using a Wesenheit function tied solely to the \citet{be07} data.   The resulting distances are 1.9 kpc, 1.8 kpc, and 1.4 kpc for V340 Nor, QZ Nor (first overtone), and GU Nor (fundamental) accordingly (formal uncertainties are on the order of $\pm0.1$ kpc).  Photometry used to compute those distances was taken from the compilation of \citet{be00} and \citet{be08}.  

\begin{figure*}[!t]
\begin{center}
\includegraphics[width=17.2cm]{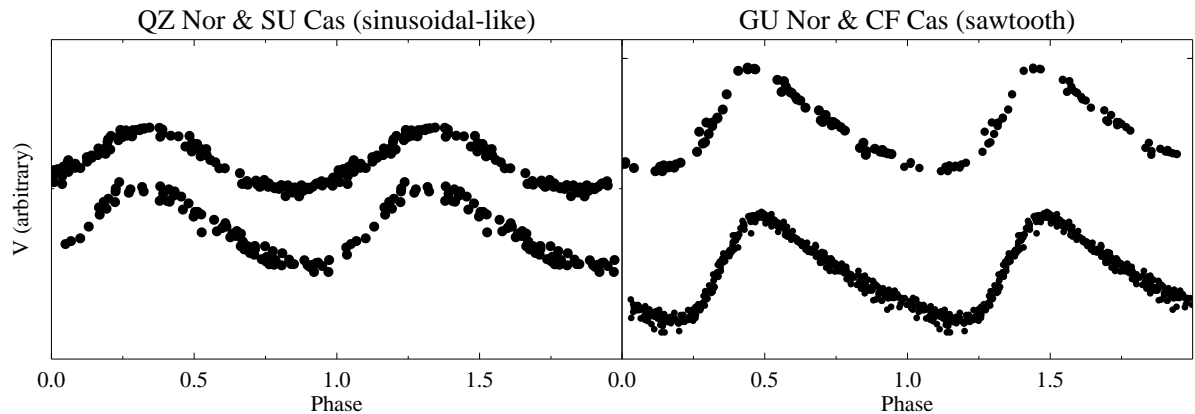} 
\caption{\small{Left, lightcurves for the sinusoidal-like Cepheids QZ Nor (top) and SU Cas (bottom), which typically imply overtone pulsation.  Right, GU Nor (top) and CF Cas (bottom) exhibit sawtooth shaped lightcurves, which are indicative of fundamental mode pulsation.  Cluster membership for SU Cas (Alessi 95) and CF Cas (NGC 7790) constrains them as overtone and fundamental mode pulsators accordingly.  The velocity evidence (Figs.~\ref{fig-rvc} and \ref{fig-rvp}) implies that QZ Nor and GU Nor are overtone and fundamental mode pulsators, respectively.  Photometry for QZ Nor, GU Nor, and CF Cas are from \citet{be00} and \citet{be08}, while the SU Cas observations were acquired via the Abbey-Ridge Observatory \citep{la08,ma08b}.}}
\label{fig-lc}
\end{center}
\end{figure*}

The Wesenheit distances are sensitive to the pulsation mode adopted.  V340 Nor ($P\sim11^{\rm d}$) is a fundamental mode pulsator, since overtone pulsators terminate near $7^{\rm d}$ in the Magellanic Clouds \citep[][see the latter's Figs.~5 \& 6]{we95,so08}, and the longest-period overtone pulsator \textit{known} in the Galaxy may be V440 Per \citep[$7.57^{\rm d}$,][]{ba09}.  For QZ Nor ($P\sim4^{\rm d}$) the matter is more complicated.  The lightcurve describing QZ Nor is sinusoidal-like (Fig.~\ref{fig-lc}), which is often argued to be a signature of overtone pulsation \citep[][and discussion therein]{sl81,gi82}. SU Cas also features a sinusoidal-like lightcurve (Fig.~\ref{fig-lc}), and is constrained to be an overtone pulsator via its membership in Alessi 95 \citep[][see also \citealt{gi76,gi82}]{tu12,ma12}.  Indeed, the \textit{majority} of Magellanic Cloud Cepheids exhibiting sinusoidal-like lightcurves occupy the Wesenheit locus tied to overtone pulsators.  However, there are instances (owing to photometric contamination, peculiarities, or otherwise) whereby sinusoidal-like Magellanic Cloud Cepheids lie on the Wesenheit ridge tied to fundamental mode Cepheids.  There exists also the cases of the enigmatic Galactic Cepheids FF Aql and Polaris, which some argue to be fundamental mode pulsators despite displaying sinusoidal-like lightcurves.  The \citet{be07} HST parallax implies that FF Aql is a fundamental mode pulsator, while the Hipparcos parallax favors overtone pulsation \citep{vl07}.  Similarly, \citet{tu13} argue on the basis of spectroscopic line ratios that Polaris' $\log{g}$ value is indicative of fundamental mode pulsation, whereas the Hipparcos parallax supports overtone pulsation \citep{vl13}.  However, the velocity evidence (Fig.~\ref{fig-rvc}) and velocity-distance gradient (Fig.~\ref{fig-rvp}) constrain QZ Nor as an overtone pulsator, given its similar velocity to V340 Nor and NGC 6067. In \S \ref{s-parameters}, it is shown that new multiband photometry implies a distance for NGC 6067 of $1.75\pm0.10$ kpc, which matches the aforementioned (first-order) Wesenheit distances.

\subsection{{\rm \footnotesize THE EXTENT OF NGC 6067}}
QZ Nor lies $20\arcmin$ from the center of NGC 6067, and that separation has helped fuel concerns toward adopting the Cepheid as a calibrator.  The cluster's extent was evaluated as follows.  2MASS data were used for the analysis since the cluster's brighter B-type stars, which rise prominently above the field in the color-magnitude diagram, are saturated in the deeper $VVV$ data.  The B-stars emerge over the field population partly because of the IMF (initial mass function), which dictates that such stars are rare relative to later-type stars.  Later-type cluster dwarfs are typically more difficult to separate from field stars.  Thus to construct the radial profile, and increase the contrast with the background population, only earlier B-type stars were used.  $BV$ data were not examined since that photometry is restricted to a smaller field of view, whereas 2MASS is all-sky.  

\begin{figure*}[!t]
\begin{center}
\includegraphics[width=14cm]{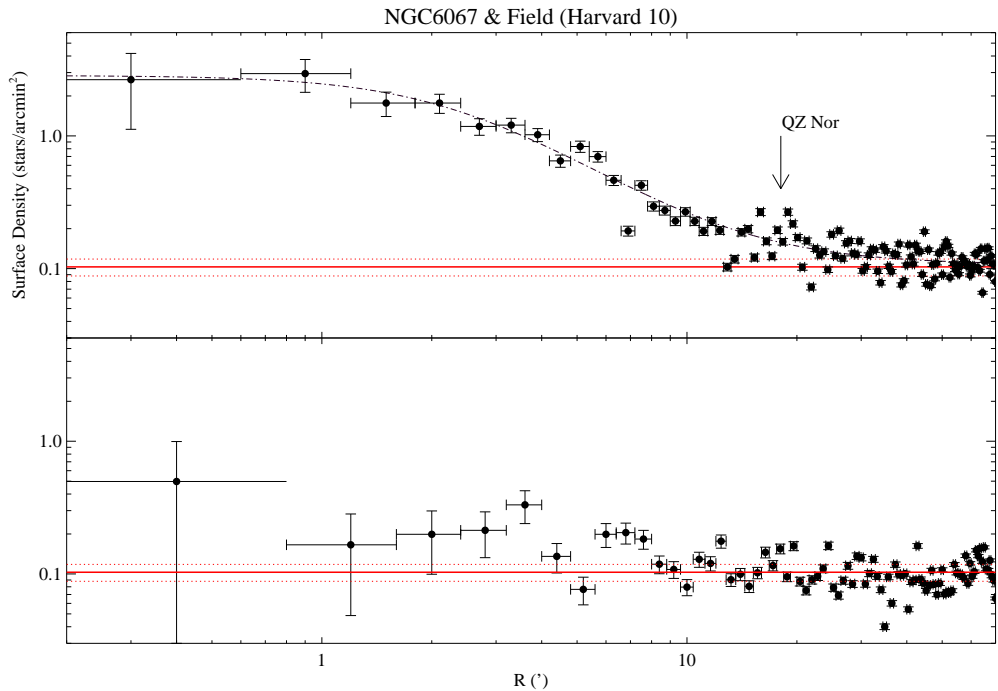} 
\caption{\small{Top panel, radial profile constructed for NGC 6067 using 2MASS observations, whereby the radial separation (arcminutes) is plotted on the x-axis. V340 Nor is positioned near cluster center, while QZ Nor lies within the cluster's corona.  GU Nor is located near the right edge of the plot.  Bottom panel, a comparison field (Harvard 10) underscores the extent of NGC 6067.}}
\label{fig-r}
\end{center}
\end{figure*}

An analysis of the 2MASS data reveals that the cluster exhibits both a sizable demographic and extent (Fig.~\ref{fig-r}).  The richness of the cluster ensures that the statistics are sufficient to extract solid conclusions.  At the position of QZ Nor, some $20\arcmin$ from cluster center, the population of NGC 6067 is larger than the background.  The counts do not level-off sharply, which may indicate the breadth of the cluster's coronal region \citep{kh69} and its dissolution into the field.  Most star clusters dissolve into the field relatively rapidly, and don't survive well beyond $10$ Myr \citep[e.g., disruption by Galactic tides,][]{ll03,bb11}.  Clusters reaching the age of NGC 6067 are thus rare.  A relatively nearby comparison field\footnote{Harvard 10 \citep[J2000 16:18:48 -54:56:00,][]{di02}.} is likewise plotted in Fig.~\ref{fig-r}, which bolsters the results established for NGC 6067.  Additional studies researching the coronal extent of NGC 6067 and similar clusters are warranted \citep{ni02,da12}.  For present purposes the conclusion is that QZ Nor's offset from the cluster center does not rule out membership.  QZ Nor lies within the bounds of NGC 6067.

\subsection{{\rm \footnotesize FUNDAMENTAL PARAMETERS FOR NGC 6067}}
\label{s-parameters}
Distance estimates for NGC 6067 exhibit a sizable spread, a partial account of which is provided below.  \citet{th62} highlight estimates dating as far back as the 1930s: 0.64 kpc (Trumpler), 0.95 kpc (Collinder), 0.751 kpc (Wallenquist), and 1.91 kpc (Shapley) \citep[see][for the references]{th62}.  \citet{th62} determined the distance to NGC 6067 based on ZAMS fitting, kinematic evidence, and spectroscopic parallaxes.  An average distance computed from the three methods yields 2.1 kpc, which is near the estimate advocated by \citet[][1.8 kpc]{en66}.  Conversely, \citet{eg83} argued that NGC 6067 was 1.3 kpc distant.  A convincing estimate for the distance to NGC 6067 was determined by \citet{wa85}.  \citet{wa85} provided the deepest $BV$ color-magnitude diagram, from which a cluster distance of 1.6 kpc was inferred.  More recently, \citet{tu10} cites a cluster distance of 1.7 kpc based partly on 2MASS, which agrees with the distance for V340 Nor established by \citet{st11} via the IRSB technique.

\subsubsection{{\rm \footnotesize NGC 6067: PHOTOMETRY}}
New $BVJH$ photometry were obtained to corroborate the \citet{wa85} determination.   $BV$ photometry was acquired via the 2.5m du Pont telescope, while the $JH$ data are from the $VVV$ survey.  The former are outlined in \citet{st09}, and further details will be restricted to a subsequent work.   The near-infrared $VVV$ data stem from PSF reductions described by \citet{mb11} and \citet{mau13} \citep[see][for the aperture photometry]{so13}.   Infrared photometry is desirable since dust extinction is less onerous than for optical observations ($A_{\lambda} \propto \lambda^{-\beta}$).  Infrared photometry thus enables deeper surveys of the obscured Galactic disk, which has resulted in the discovery of numerous star clusters \citep{bo11,ch12,ma12c}.  The impact of variations in the extinction law is also mitigated in the near-infrared, given $E(J-H)\sim0.3 \times E(B-V)$ and $R_J\sim2.7$ \citep[][and references therein]{bo04,ma11a}. However, employing multiband (optical+infrared) photometry is most desirable to reduce the influence of systematic errors.  A comparison of a cluster's distance inferred from optical/infrared photometry can subsequently ensue \citep[e.g., Pismis 19,][]{ca11,ma12e}.

\subsubsection{{\rm \footnotesize NGC 6067: REDDENING}}
Spectral types for stars in NGC 6067 \citep{th62} are tabulated in WEBDA \citep{mp03} and \citet{sk13}.  The resulting mean color-excess for 13 cluster stars with spectral types designated by \citet{th62} is $E(B-V)=0.35\pm 0.04 (\sigma)$.  Intrinsic $(B-V)_0$ colors were adopted from \citet[][and references therein]{tu89}.  The resulting near-infrared excess is $E(J-H)\sim0.12$, assuming $E(J-H)\sim0.33 \times E(B-V)$. The reddening is $E(J-H)\sim0.14$ when using intrinsic $(J-H)_0$ colors from \citet{sl09}, together with the aforementioned \citet{th62} stars, and stars from \citet{sk13}.  A mean value of $E(J-H)=0.13$ was adopted.  

The reddening and extinction laws appear to vary across the Galaxy \citep{ca12,na12}. The extinction law ($R_V$) for $\ell\sim330\degr$ lies near the canonical value of $R_V\sim3.1$, as inferred from the work of \citet{fm07} and Majaess et al.~(2013-14, in preparation).  The latter applied the color ratio extrapolation method to mid-infrared WISE-Spitzer data.  

\subsubsection{{\rm \footnotesize NGC 6067: DISTANCE}}
\label{s-age}
To determine the cluster distance an isochrone was shifted in magnitude space to match the observed data, since the color-excess was determined above and abundance estimates for the cluster Cepheids are near solar \citep{lu11}.  Padova isochrones were used once zero-pointed to the distance scale of \citet{ma11c}.  An empirical $JHK_s$ main-sequence calibration was established by \citet{ma11c} using deep 2MASS photometry and revised Hipparcos parallaxes \citep{vle07} for nearby stars ($d<25$ pc).  The infrared calibration is comparatively insensitive to stellar age and [Fe/H], and is anchored to seven benchmark open clusters that exhibit matching $JHK_s$ and revised Hipparcos distances \citep{vl09,ma11c}.  The objective was to avoid deriving distances to Cepheid clusters using a single benchmark cluster (i.e., the Pleiades), and thus introduce a potential systematic uncertainty into the Cepheid calibration since the Pleiades distance is contested \citep[][and references therein]{vl09,ma11c,dg12}.  A visual fit yields a near-infrared distance of $d=1.75\pm0.10$ kpc for NGC 6067, and the optical data provide an analogous result to within the uncertainties.  The $VVV$ data were supplemented by 2MASS observations at the bright end.  Red giants sharing the Cepheids' velocities \citep{me08} were likewise added to the color-magnitude diagrams.  The isochrone fit
and uncertainties were established via the traditional visual approach \citep[e.g.,][]{cm04,bb10}, and the latter represents the limit where a mismatch is clearly perceived. \citet[][and references therein]{pn06} note that errors tied to isochrone fitting via computer algorithms are comparable to those associated with the traditional approach. The deep photometry provided reliable anchor points to facilitate the isochrone fitting. The present result is smaller than the cluster distance advocated by \citet[][$d \simeq 2.1\pm0.3$ kpc]{th62}, and in closer agreement with the \citet{wa85} result ($d=1.62\pm0.07$ kpc).  An age of $\log{\tau}=7.90\pm0.15$ appears to match the cluster's B-type and red giant members.  

\begin{figure*}[!t]
\begin{center}
\includegraphics[width=11cm]{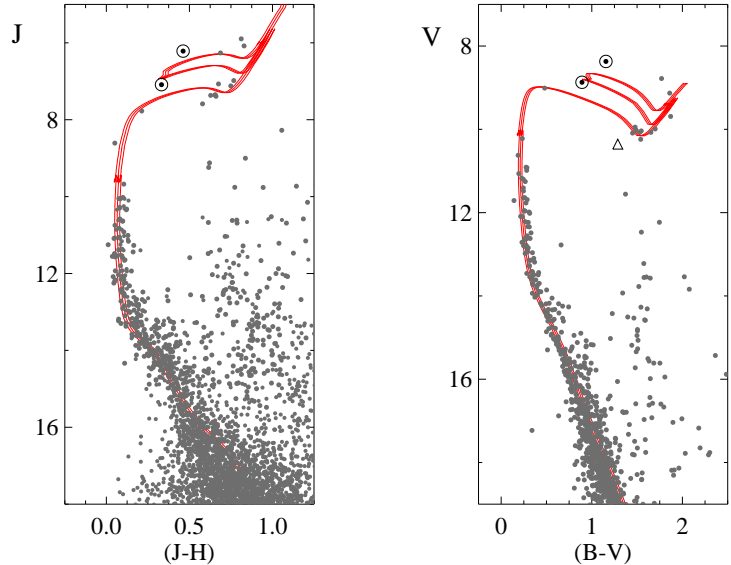} 
\caption{\small{New $BVJH$ photometry was used to constrain the distance for NGC 6067.  A Padova isochrone fit to cluster stars yields $d=1.75\pm0.10$ kpc and $\log{\tau}=7.90\pm0.15$, once shifted by the mean color-excess.  The circled dots represent QZ Nor and V340 Nor (brighter object), while the open triangle represents GU Nor.  Left, near-infrared color-magnitude diagram constructed using $VVV$/2MASS photometry, while the plot on the right features optical data from the du Pont telescope and \citet{me08}.}}
\label{fig-cmd}
\end{center}
\end{figure*}

\begin{deluxetable}{lcccc}
\tablewidth{0pt}
\tabletypesize{\small}
\tablecaption{Estimated Ages for the Cepheids (Myr)}
\tablehead{\colhead{ID} & \colhead{$\tau$ (E03)\tablenotemark{1}} & \colhead{$\tau$ (T12)\tablenotemark{1}} & \colhead{$\tau$ (B05)\tablenotemark{2}} & \colhead{$\tau$ (B05)\tablenotemark{2}}}
\startdata
QZ Nor & 106 & 91 & 84 & 74 \\
V340 Nor & 65 & 52 & 40 & 51 \\
GU Nor & 141 & 126 & $...$ & $...$  \\
\enddata
\tablenotetext{1}{\scriptsize{Results from the period-age relations of \citet[][E03]{ef03} and \citet[][T12]{tu12b}.  The standard deviation deduced from all age estimates for QZ Nor and V340 Nor is $\sigma_{\tau}\sim12$ Myr \citep[see also Figs.~3 and 8 in][]{bo05}.}}
\tablenotetext{2}{\scriptsize{The model-based results from \citet[][B05]{bo05} were copied verbatim from their Table 9.  However, a minor typographical error appears to exist with the root expressions in their Table 6, as the final values in Table 9 were not reproducible.  Thus no ages are cited for GU Nor granted the star was not analyzed in that study (however see Figs.~3 and 8 in B05).   The period-age and period-age-color results deduced by B05 for QZ Nor and V340 Nor are stated in the $4^{\rm th}$ and $5^{\rm th}$ columns of the present table, respectively.}}
\label{table1}
\end{deluxetable}

NGC 6067 is not the only cluster to host multiple Cepheids.  For example, three Cepheids are members of NGC 7790 \citep{pe84,ma95}, and \citet{ef03} noted that NGC 1958 (LMC) features 6-9$^{\rm d}$ Cepheids.  Yet NGC 6067 exhibits the largest period spread among its Cepheid constituents.  The $\Delta P_0 \sim 6^{\rm d}$ offset between QZ Nor and V340 Nor is admittedly concerning since a period-age relation exists \citep{ef03,tu12}. However, \citet{bo05} argued that the age offset between QZ Nor and V340 Nor could be mitigated by using a period-age-color relation derived from models, and the reader is referred to that study for details. \citet{bo05} concluded that the age spread is reduced from 44 Myr (period-age) to 23 Myr (period-age-color).  The period difference between GU Nor ($P\sim3^{\rm d}.5$) and V340 Nor is sizable ($8^{\rm d}$). The period-age relationships of \citet{ef03} and \citet{tu12} imply that GU Nor is $\tau \sim 134$ Myr (Table~\ref{table1}), which is older than the cluster and other Cepheids (e.g., V340 Nor).  \textit{Taken as a whole}, the suite of evidence suggests that GU Nor is a member of the field population and not bound to NGC 6067.   

\section{{\rm \footnotesize CONCLUSION AND FUTURE RESEARCH}}
\label{s-conclusion}
A multi-faceted approach undertaken implies that V340 Nor and QZ Nor are members of NGC 6067, whereas GU Nor is likely a foreground star.  Radial velocities for the two former Cepheids and cluster agree within 1 km/s, whereas GU Nor is discrepant (Fig.~\ref{fig-rvc}).  The predicted velocity-distance correlation inferred from Galactic rotation yields a steep gradient for the NGC 6067 sight-line ($\ell \sim 330 \degr$, Fig.~\ref{fig-rvp}), indicating that radial velocities are a reliable (relative) distance indicator.  Hence, objects sharing common velocities are nearly equidistant (i.e., QZ Nor/V340 Nor/NGC 6067, Figs.~\ref{fig-rvc}, \ref{fig-rvp}).  The radial extent of NGC 6067 was mapped using 2MASS (Fig.~\ref{fig-r}), and it was demonstrated that QZ Nor lies within the cluster boundary, particularly when the cluster's  corona and dissolution into the field are considered.  New $BVJH$ photometry was employed to determine a precise distance to NGC 6067 ($d=1.75\pm0.10$ kpc), which matches Wesenheit distances computed for QZ Nor and V340 Nor using the \citet{be07} calibration (\S \ref{s-rvs}).  In sum, the conclusions derived here support prior assertions by \citet{wa85} and \citet{cc85} concerning the membership status of the aforementioned Cepheids.   QZ Nor is a cluster Cepheid that can help anchor the distance scale.   GU Nor is likely a member of the field population and not bound to NGC 6067, as indicated by the Cepheid's radial velocity, Wesenheit distance, and age.

The present analysis would benefit from new high-resolution spectra for numerous cluster members.  Those spectra would permit a more reliable determination of the dust properties and cluster color-excess.  Furthermore, spectra and $UBV$ photometry are needed for bright B-stars in close proximity to QZ Nor, in order to determine that Cepheid's reddening. Mean reddening estimates compiled for V340 Nor and QZ Nor by \citet{fo07} imply that there is a $\Delta E(B-V)\sim0.07$ differential offset between the Cepheids, a result that would benefit from independent confirmation.   Assessing the period evolution of the Cepheids will likewise be beneficial \citep{tu06,ne12b}, as the analysis can independently confirm that V340 Nor and QZ Nor lie in separate crossing modes (Fig.~\ref{fig-cmd}).  Preliminary efforts to ascertain the period evolution of QZ Nor/V340 Nor were marred by a short temporal baseline \citep[the Cepheids were discovered relatively recently,][]{eg83}.  A sizable temporal baseline is required to separate a Cepheid's long-term secular changes in period owing to stellar evolution, from short-term variations stemming from binarity or random and unknown behavior.  The Harvard College Observatory houses photographic plates featuring QZ Nor that date back to the 19$^{\rm th}$ century.  Visual estimates from those plates will be needed unless surveyed by DASCH, which is a project aimed at digitizing the Harvard collection \citep{gr12}.  However, QZ Nor and V340 Nor are small-amplitude Cepheids \citep{ks09}, which will exacerbate the uncertainties.

Lastly, despite recent gains in bolstering the short and intermediate period regimes of the Galactic Cepheid calibration, considerable effort remains to constrain the long-period domain.   A well-sampled calibration featuring long-period Cepheids is desirable, as remote extragalactic Cepheids observed using HST/LBT typically exhibit periods greater than 10 days \citep{ss11,ge11}, as their shorter-period counterparts are fainter.  A well-sampled long-period cluster Cepheid calibration would likewise foster stronger constraints on the $p$-factor, which is used to establish IRSB distances \citep{gi05b,st11,ng12,jl12,ne12b}.  A multi-object fiber-fed spectrograph can survey numerous stars surrounding long-period Cepheids, and automated methods can classify the resulting sample and establish spectroscopic parallaxes \citep{man09,mr13}.  Those parallaxes could be examined for overdensities, and the resulting distance and age of the constituents compared to first-order predictions for the target Cepheid.  \citet{ma11b} and \citet{ma12b} also highlighted approaches by which the long-period end of the Galactic calibration could be confirmed.  The former noted that long-period Cepheids in the Galaxy's spiral arms could be used to calibrate Leavitt's law, while the latter demonstrated that X-ray observations (e.g., XMM-Newton ID 0603740501, PI Guinan) are helpful for establishing precise distances to nearby clusters hosting Cepheids\footnote{Recent observations imply that Cepheids are X-ray (low flux) emitters \citep{eg12}.} (e.g., Alessi 95).  X-ray observations are pertinent for such efforts since they facilitate the identification of stars associated with Cepheids \citep{ev11}. Younger stars linked to Cepheids can be segregated from field stars along the sight-line, which are typically old slow-rotators that have become comparatively X-ray quiet.   The reliability of distances established to Cepheid clusters, based on fitting evolutionary tracks to the color-magnitude diagram, is partly commensurate with the number of cluster stars identified.  Obtaining X-ray data for comparatively nearby longer-period Cepheids is desirable.  

\subsection*{Acknowledgements}
\scriptsize{DM is grateful to the following individuals and consortia whose efforts, advice, or encouragement enabled the research: Y. Efremov, H. Neilson, P. Moskalik, 2MASS (R. Cutri), A. Thackeray, A. Walker, I. Coulson, J. Caldwell, F. Kienzle, M. Metzger, M. Groenewegen, D. Bersier, F. Pont, J-C. Mermilliod, HIP (F. van Leeuwen), HST (F. Benedict, B. McArthur), OGLE (A. Udalski, I. Soszy{\~n}ski), G. Bono, B. Skiff, D. Balam, WEBDA (E. Paunzen, J-C. Mermilliod), W. Dias, Spitzer, WISE, CDS (F. Ochsenbein, T. Boch, P. Fernique), arXiv, and NASA ADS.  WG, DG, and D. Minniti are grateful for support from the BASAL Centro de Astrofisica y Tecnologias Afines (CATA) PFB-06/2007.}

\end{document}